# Optimizing musical chord inversions using the cartesian coordinate system

Steve Mathew D A*
*Vellore Institute of Technology, Vellore, India
*Indian Institute of Technology Madras, Chennai, India

**Abstract**

In classical music and in any genre of contemporary music, the tonal elements or notes used for playing are the same. The numerous possibilities of chords for a given instance in a piece make the playing, in general, very intricate, and advanced. The theory sounds quite trivial, yet the application has vast options, each leading to inarguably different outcomes, characterized by scientific and musical principles. Chords and their importance are self-explanatory. A chord is a bunch of notes played together. As far as scientists are concerned, it is a set of tonal frequencies ringing together resulting in a consonant/dissonant sound. It is well-known that the notes of a chord can be rearranged to come up with various voicings (1) of the same chord which enables a composer/player to choose the most optimal one to convey the emotion they wish to convey. Though there are numerous possibilities, it is scientific to think that there is just one appropriate voicing for a particular situation of tonal movements. In this study, we attempt to find the optimal voicings by considering chords to be points in a 3-dimensional cartesian coordinate system and further the fundamental understanding of mathematics in music theory.

Email: stevemathewda@gmail.com

# 1 Introduction

The concept of voicings and inversions are at many times overseen, and they aren't given the importance they deserve. It is worthwhile to note that the voicings are not always meant to be specified in sheet music. Yet to quote the Grammy Winning Composer and Musician Jacob Collier, there can be a lot of emotions conveyed just with the 3 different voicings of a basic major triad. Though it is a perceptual statement, it is cemented by a basic, fundamental property of sound propagation and tonal frequencies that suggests that the tonic triad deserves the most stable note in the scale as its high note.

One basic understanding that has always been the fundamental rule for choosing the inversion is the need to maintain the lowest tonal tension possible. It can be attained by reducing the tonal distance travelled in a measure of music. Having all the chords near to each other, in other words, demanding the least finger movements in the case of a piano, considerably reduces the fluctuations of frequencies instead of approaching unnatural highs and lows. This is a very important concept, and it perfects the utmost cause of playing chords; to sound pleasing. Making advancements in the same topic using the above perspective by considering chords to be points in a 3-dimensional space, we arrive at 3 different set of inversions with respect to the same major scale (Table. 1) or minor scale (Table. 4). It is worthwhile to note that the inversion aka. voicing which has the tonic as the first note is the 1st inversion, the major 3rd as first note is the 2nd inversion and the dominant as the first note is the 3rd inversion. The entries in the table refer to degrees of notes that are with respect to the chord built upon the tonic in discussion. By following any of the 3 sets of inversions, one can be assured that they're play has a minimal amount of tension and that the transitions (musically, progressions) are smooth.

# 2 Methods

## 2.1 Formation of inversion sets maintaining minimal tension

Step 1: Choose the tonic and start from the root position of the tonic triad.
Step 2: Try figuring out various inversions of other diatonic chords in the scale facilitating minimal movement in between chords while moving from the tonic triad.
Step 3: Congruency is key. Here, congruency refers to the ability to maintain the same inversions throughout the piece unless the melody leads the song to either of the extremities – bass/treble tones.

## 2.2 Notion of perceptual analysis in voicing

Perceptually qualifying chord voicings is regarded as one of the easiest ways to finding the optimal voicings. In most of the times, this is thought to be enough as far as the practical application is concerned. Scientific implications can be quite intriguing and might not go in agreement with the conventional styles of playing. Yet, the implications are framed by mathematical proofs corroborating the fundamentals of sonic phenomena. The disadvantage of perceptual analysis is that human minds tend to stick onto one set of standard inversions (5,6,7) and might not embrace the other options even when variation in inversions gives a completely distinct, pleasing feel to the piece.

## 2.3 Reasons for choosing voicings mathematically

It is to be remembered that the frequencies of piano notes (or any sequence of musical notes for that matter of fact), follow an exponential increase across the board from bass to treble notes (2). It is obvious that the higher frequencies (shrieking ones) have a sort of 'ring' when they propagate, making the sound more evident. This ring is attributed to the fact that the difference in the frequencies of consecutive notes in bass end is less and the difference in the frequencies of consecutive notes in the treble end is high (3) which even helps us characterize the frequencies of the piano using a single exponential equation. The same concept of high notes being strongly evident to the human hearing in chords applies as, the note that is located on the farthest right on the chord or that which technically has the highest frequency would be coming out the best. Thus, the note that is about to ring out has a greater impact than that of the other two notes in the triad.

It is to be noted that every chord that is diatonic to a particular scale is graphically 'in sync' with that of the tonic (3,4). Yet the degree of consonance varies with each note. That is the tail of the problem that gives us an edge to arrive at a concrete conclusion. It is obvious that the first preference of the inversion would be to have the unison as the highest note in the tonic triad. Which makes the second column of inversions of Table 1, the most consonant. Following the same, it would be the first column which will be the 2nd most consonant group as it has the dominant as the higher note which is the second most consonant note in the scale (3,8). It is to be noted that Table 2 deals with the diatonic chords of a major scale. Extrapolating the same results to minor scales is elucidated further down the study.

## 2.4    Tonic Triad: The Nucleus

It is to be noted that the start (first note) or at least the end (last note) of many major scaled compositions is going to be the tonic major triad or the constituents of the same. Thus, it would be very sensible to have the tonic aka. key-center as the highest note in the tonic major triad. This might not prove to be helping for the other triads diatonic to the scale that might not have the tonic as the highest note though it contains it since these other diatonic chords need to maintain the lowest possible tonal movement from the tonic triad. But as far as the big picture of the proceedings of the scale is concerned, it is worthwhile to mention that the best argument for assigning the tonic as high note for tonic triad would be that every other chord in the scale is revolving around the tonic chord with their respective inversions very close to the latter which in turn is revolving around the note, the tonic which would have the mentioned 'ring' when it is located relatively higher the board. Doing this will aid to the fact that we are trying to emphasize the sound of the tonic. This is done since our mind always tries to relate the sequence of notes in a piece to its key-center (9,10).

**Table 1.** Possible sets of voicings with the least tonal movement for a major scale

| Tonic of the chord | Chord Type | First note of inversion - 1 | First note of inversion - 2 | First note of inversion – 3 |
|---|---|---|---|---|
| **Tonic** | Major | Tonic | Major 3$^{rd}$ | Dominant |
| **Supertonic** | Minor | Tonic | Minor 3$^{rd}$ | Dominant |
| **Major 3$^{rd}$** | Minor | Dominant | Tonic | Minor 3$^{rd}$ |
| **Subdominant** | Major | Dominant | Tonic | Major 3$^{rd}$ |
| **Dominant** | Major | Major 3$^{rd}$ | Dominant | Tonic |
| **Major 6$^{th}$** | Minor | Minor 3$^{rd}$ | Dominant | Tonic |
| **Leading Tone** | Diminished | Tonic | Minor 3$^{rd}$ | The flat 5$^{th}$ |

## 2.5 Plotting chords in a 3-dimensional graph to identify the inversions with the least tonal movements

The Table 1 is arrived at by assigning the values of the chromatic scale to integers starting from 0 to 23 (at least 2 octaves) as shown in Table 2

**Table 2.** Sequence of two chromatic scales starting in C

| C | C# | D | D# | E | F | F# | G | G# | A | A# | B | C | C# | D | D# | E | F | F# | G | G# | A | A# | B |
|---|---|---|---|---|---|---|---|---|---|---|---|---|---|---|---|---|---|---|---|---|---|---|---|
| **0** | 1 | 2 | 3 | **4** | 5 | 6 | **7** | 8 | 9 | 10 | 11 | **12** | 13 | 14 | 15 | 16 | 17 | 18 | 19 | 20 | 21 | 22 | 23 |

Let us conduct this analysis with respect to the C Major Scale. The same method can be used to any major scale and the same results would follow. In the above table, the notes of the C Major Chord are highlighted in a sequence such that the tonic of the triad is located at the right-most position possible. Let us consider a 3-dimensional graph on which we are attempting to plot each triad as a point with 3 co-ordinates. So here, the C major triad has the co-ordinate (4,7,12) on the space. Similarly, we are going to plot the other triads diatonic to the C Major Scale. But when doing so, we are to make sure that the distance of the other diatonic chords is the least possible distance from the C major triad which is graphically located at (4,7,12). Thus, the D minor chord which consists of the notes D, F and A must be in the second inversion (F, A and D) which will be situated

at (5,9,14) on the graph. The same is elaborated as follows:

We assume that the tonic triad's best inversion is E-G-C since the tonic is at the right-hand position. Let us consider the tonic chord to be of the form $a = (x_a, y_a, z_a)$. Let every other chord be taken as $b_i = (x_{bi}, y_{bi}, z_{bi})$ where i ∈ {1,2,3,4,5,6} since there are 6 other diatonic chords in the scale apart from the tonic triad.

Let the chord *a* be C Major with the coordinates (4,7,12). Let the chord *b* be D minor. But *b* can be situated in any one of the points of the set {(2,5,9),(5,9,14),(9,14,17)}. The chord can also be in the subsequent points, but they would be the same 3 inversions in higher octaves. When the distance between the points (4,7,12) and (5,9,14) are calculated, they turn out to be the least when compared with other combinations.

$$\text{Distance between chords } a \text{ and } b_i = \sqrt{(x_{bi} - x_a)^2 + (y_{bi} - y_a)^2 + (z_{bi} - z_a)^2}$$

When the same was done with all the diatonic chords, the inversions that had the least distance from the tonic major triad are the ones that are specified in Table 1.

Location of chord $a = (4,7,12)$

**Table 3**. Graphical distances between the tonic triad and other diatonic triads in a major scale

| Chord of interest (*b*) | First note in the inversion | Closest possible location to point *a* | Distance between *a* and *b* |
|---|---|---|---|
| D minor | Minor 3rd | (5,9,14) | 3 units |
| E minor | Tonic | (4,7,11) | 1 unit |
| F major | Tonic | (5,9,12) | 2.23606 units |
| G major | Dominant | (2,7,11) | 2.23606 units |
| A minor | Dominant | (4,9,12) | 2 units |
| B diminished | Minor 3rd | (2,5,11) | 3 units |

The total of all the distances is ~13.472 units. The interesting point to note is, the cumulative distance when calculated with a different inversion of the tonic triad (the inversions of the other diatonic chords will also change), remains to be the same 13.472.

Thus, irrespective of the note which is the highest in the tonic triad, the sum of distances between the tonic triad and the other diatonic chords remains constant. And that constant value is always 13.472 (since, even when the same method is applied to other major scales, mathematically the entire scale is only transposed, and no other change is incurred).

When we attempt to do the same for a minor scale, we get a similar table (Table. 4). Even in this case, the second column of inversions has the tonic as the treble note in the in the triad. Even for the analysis of minor scale, we take the same chromatic sequence that was used in Table 2.

**Table 4.** Possible sets of voicings with the least tonal movement for a natural minor scale

| Tonic of the Chord | Chord Type | First note of Inversion - 1 | First note of Inversion - 2 | First note of Inversion – 3 |
|---|---|---|---|---|
| **Tonic** | Minor | Tonic | Minor $3^{rd}$ | Dominant |
| **Major $2^{nd}$** | Diminished | Tonic | Minor $3^{rd}$ | The flat $5^{th}$ |
| **Minor $3^{rd}$** | Major | Dominant | Tonic | Major $3^{rd}$ |
| **Subdominant** | Minor | Dominant | Tonic | Minor $3^{rd}$ |
| **Dominant** | Minor | Minor $3^{rd}$ | Dominant | Tonic |
| **Minor $6^{th}$** | Major | Major $3^{rd}$ | Dominant | Tonic |
| **The seventh** | Major | Tonic | Major $3^{rd}$ | Dominant |

We are considering the A minor scale (the relative minor of C major) to conduct the analysis for minor scales. The following is the location of the tonic triad such that the root of the tonic triad is located as the treble note.

Location of chord $a = (12,16,21)$

**Table 5**. Graphical distances between the tonic triad and other diatonic triads in a natural minor scale

| Chord of interest (*b*) | First note of inversion | Closest possible location to point *a* | Distance between *a* and *b* |
|---|---|---|---|
| B diminished | Minor 3$^{rd}$ | (14,17,23) | 3 units |
| C major | Tonic | (12,16,19) | 2 units |
| D minor | Tonic | (14,17,21) | 2.23606 units |
| E minor | Dominant | (11,16,19) | 2.23606 units |
| F major | Dominant | (12,17,21) | 1 unit |
| G major | Major 3$^{rd}$ | (11,14,19) | 3 units |

Even in the case of minor scales, we get the same cumulative tonal distance between the tonic triad and the other diatonic chords, and the cumulative distance is a constant for all the three inversion sets (~13.47 units). This is significant because, the fundamental structure of both the major and the minor scales are contrastingly different, and yet they share the same minimal cumulative tonal distance. Thus, this ensures that the jubilancy that is associated with the major scales, or the seemingly sorrowful taste given by the minor scales are not associated in any way with the number of minimum movements that are needed at the chordal level but are rather subjected to the intervals using which the scales are constructed.

## Conclusional Remarks

This study dealt with trying to understand the inversion sets of diatonic chords that have the least possible tonal movement. To facilitate this, we considered chords to be points on a 3-dimensional space and achieving the smallest possible cumulative tonal distance was the goal when the appropriate positions for the other diatonic chords in the scale were found. This moved us towards the next finding which is that the minimal tonal distance from the tonic triad to other diatonic chords is irrespective of whether the scale is major or minor and the distance is always a constant (~13.47 units). This is a substantial proof that the mood rendered by the major and minor scales are only attributed to the constituent notes of the scales and not to any other consonant or dissonant feel that could be an effect of how the diatonic chords are arranged in the scale. The inversion sets gained from the analyses give us three options to choose from (depending on which inversion of the tonic triad we are going to use) and the inversions of the other diatonic chords change depending on which set we chose. Yet again, mathematical analyses proved to be effective in enhancing music performance by helping us choose the inversions of the diatonic chords in a song, keeping the tonal movements to the possible minimum.

# References

1. Catherine Schmidt-Jones, 2015, Understanding Basic Music Theory, ISBN Number: 9781508534297
2. David Wright, 2009, Mathematics and Music. ISBN Number: 9780821848739
3. Mathew, Steve. "Tonal Frequencies, Consonance, Dissonance: A Math-Bio Intersection." arXiv preprint arXiv:2106.08479 (2021).
4. Mathew, Steve (2021): The Resolution of Musical Chords: A Physical Paradigm. Advance. Preprint. https://doi.org/10.31124/advance.14774928.v1
5. Ohio State University. "This is your brain detecting patterns: It is different from other kinds of learning, study shows." 31 May 2018, ScienceDaily.
6. CNRS (Délégation Paris Michel-Ange). "How does the human brain memorize a sound?", 2 June 2010, ScienceDaily.
7. Josh McDermott et al, "Understanding how the brain makes sense of sound" – Massachusetts Institute of Technology – National Science Foundation (2019)
8. Terhardt E (1974) Pitch, consonance, and harmony. J Acoust Soc Am 55: 1061-1069.
9. Deutsch D (1973) Octave generalization of specific interference effects in memory for tonal pitch. Percept Psychophys 13: 271-275.
10. Carol L. Krumhansi (2001) Cognitive Foundations of Musical Pitch, Oxford Psychology Series, Oxford University Press; ISBN Number: 0190287446, 9780190287443